\documentclass[11pt,a4paper]{article}
\usepackage{graphicx}
\usepackage[utf8]{inputenc}
\usepackage[english]{babel}
\usepackage{amsmath,amssymb} 
\usepackage{color}
\usepackage{float}
\usepackage{hyperref}
\usepackage{pdfpages}
\usepackage[square,numbers,comma,sort&compress]{natbib}

\begin{document}

\title{Representing pictures with emotions} 

\author{António Filipe Fonseca \footnote{\href{mailto:ajffa@iscte-iul.pt}{ajffa@iscte-iul.pt} \newline ISTAR/ISCTE-IUL Av. das Forças Armadas 1649-026 Lisboa Portugal} }

\date{Received: 06/12/2018}

\maketitle

\begin{abstract}
Modern research in content-based image retrieval systems (CIBR) has become progressively more focused on the richness of human semantics. Several approaches may be used to reduced the 'semantic gap' between the high-level human experience and the low level visual features of pictures. Object ontology, among others, is one of the methods. In this paper we investigate the use of a codified emotion ontology over global color features of images to annotate the images at a high semantic level. In order to speed up the annotation process the images are sampled so that each digital image is represented by a random subset of its content. We test within controlled conditions how this random subset may represent the adequate high level emotional concept presented in the image. We monitor this information reducing process with entropy measures, showing that controlled random sampling can capture with significant relevance high level concepts for picture representation.
\end{abstract}

\section{Introduction}
\label{intro}
The development of the internet and of computer in general is bringing each day to the millions of users a more humanized and intelligent experience. Modern technology of digital information retrieval progressively provide less stressing, quicker and more relevant experience in the search and in the retrieval of information. In the context of content-based image retrieval systems (CIBR) the recent trends show that the focus of research in is being shifted from designing sophisticated low-level feature extraction algorithms to reducing the 'semantic gap' between the visual features and the richness of human semantics \cite{YingLiu07}. This 'semantic gap' may be narrowed through the use of different methodologies: (1) using object ontology (2) using machine learning techniques (3) using relevance feedback from the users (4) generating semantic templates from low level features (5) associating evidences from HTML and visual content for WWW image retrieval \cite{YingLiu07}.\\

In some cases, semantics can be easily derived from our daily language, turning the application of an object ontology into straightforward process. For example, sky can be described as (upper, uniform, smooth texture and blue region) of a picture, or grass can be described as (lower, uniform, rugged texture green region) in the picture. Although features like position and segmentation are relatively easy to detect and qualify in an image, texture is less prow to be modeled and understood, so color features are instead widely used as can easily be associated with large classes of object.\\

The use of color as a component in ontology based system requires the quantization and naming. An equivalent to the naming process is the direct association of color with emotion related semantics like human feelings or ambiences. Although loosing precision when ignoring shape, texture, position or other features of objects, using color and its emotional traits has the great advantage of broader recall while narrowing the semantic gap. This process can be further complemented with other search criteria. Several researchers already propose different color based semantic classification methodologies for images.\\

In this paper we are not as much concerned with the methodology in color classification as with the problem of performing this classification with reduced information over each item in the dataset. The result of examining only a fraction of the information available in each image and still be able to aport relevant features into the classification process is our major concern. The problem of reduced information over the available information in the original image may be seen from three distinct perspectives: (1) truncation, (2) noise and (3) quantization noise.\\

The problem of removing noise from images, and the more general problem of reconstructing a signal that has been corrupted in some way is by itself a autonomous area of research. Although quantization noise is already implicit in color quantization or in any other low-level classification of the image, we will be more concerned here with the more restrict problem of random truncation of the whole content: the problem of obtaining semantic relevant classification in the presence of a sub-sampling of the original image. Moreover we will be concerned with the advantage of sub sampling the image in order to speed up the classification process.\\

In section 2 we will revise related work about color and emotional annotation and classification of images. In section 3 we will present a methodology of emotional classification based in low-level color features. In section 4 we will examine the inherent problems with the coding of the color information in the classified image. In section 5 we will approach the problem of sampling the full extent of the image. Finally in section 6 we will present our discussion and conclusions.

\section{Related Work}
\label{sec:relwork}
The use of emotion and moods as annotation of digital objects is not novel and has already been studied not only to classify pictures but also sound and movie pieces. Riad et al. \cite{Riad12} about automatic annotation of images distinguish two types of annotation: keywords-based annotation and ontology-based annotation. The second type being more versatile to the general user with low capability to analyze semantic relations among keywords. In our work we will use color to annotate the emotions and moods salient in each picture. In this sense we are supporting our classification in a complex many to many correspondence between colors and emotions. Csurka et al. \cite{Csurka10} propose a complex vector color-space correspondence between colors and abstract categories. Other authors \cite{Solli10} use color scales like Shigenobu Kobayashi's \textit{Color Image Scale} \cite{Kobayahi91} or the LHC color space \cite{Wang05} which better corresponds to human perception, in \cite{Wang06} the authors use an orthogonal three-dimension emotional factor space with 12 pairs of emotional words. Related to emotion is the concept of harmony between colors, Cohen-Or \cite{Cohen-Or06} use the properties of color hue scale to implement an algorithm for color harmonizing.\\

In our work we employ a reference work for web design \cite{Morton97} which has the advantage of cross classification of emotions and moods between colors within a large palete of 108 colors. The book by Jill Morton establishes a many to many correspondence between a 108 color palete and a vocabulary of 187 different moods and sensations.\\

Although some authors use machine techniques to implement the annotation process \cite{dunker08} \cite{Csurka10} others support its algorithms on other statistical mappings between low-level features of images and the high-level emotional content \cite{Solli09} \cite{Solli10} \cite{Machajdik10}. Solli \cite{Solli09} implements a complex process of conversion between scale-space representations of emotion channels and aggregates of emotional feeling within the image. In \cite{Solli10} the author uses only hue and tone representations of the image to map the emotions. In \cite{Machajdik10} the authors employ assisted learning to classify diferent kinds of low-level image features like color, texture and composition with the aid of pre-classified image datasets. Qingyong Li \cite{Li2007}, implements the classification based in direct a fuzzy representation of color in the image, whereas in \cite{Agnieska10} a neural network is trained in order to map low-level features into four groups of adjectives, five basic emotions and a positive/negative tag.\\

In our work, given the richness of the color-emotion dataset, we chose to map emotions directly from the image pixels with further statistical processing which proved to be a robust process. Our purpose was to devise a method able to support the random truncation of information belonging to the original image.

\section{Emotion Processing}
\label{sec:emopro}
In order to associate emotional adjectives with the color of the images we used a emotion palette os colors \cite{Morton97} which makes a correspondence between colors and attributes of psychological symbolism. This correspondence is not unique and may depend on several factors of which the most important is culture. This issue constitutes by itself an whole area of research. The attributes are psychological related but colors could be associated differently, to natural references like \textit{sunlight}, \textit{sand}, \textit{fish} or \textit{snow} or to contemporary culture like \textit{taxi} for yellow or \textit{chocolate} for brown, although these alternatives would narrow the range of applicability of the palette.\\

\begin{table}[t]
\begin{center}
\label{tab:table1}
\caption{Examples of color annotation}
\vspace{0.5cm}
\begin{tabular}{|l|l|}
	\hline\noalign{\smallskip}
	color & emotions\\
  \hline \noalign{\smallskip}
  Dark Brick Red & earthy, friendly, robust, strong, tasty, warm \\
  Salmon Red & healthy, happy, tasty, friendly, cosmetic, warm \\
  Deep Plum Red & elegant, majestic, spiritual, fruity, feminine \\
  Medium Burgundy Purple & vibrant, spiritual, passionate, floral, fruity\\
  \dots & \dots\\
  \hline\noalign{\smallskip}
\end{tabular}
\end{center}
\end{table}

The process of emotion annotation of the images is implemented in four steps:
\begin{itemize}
\item A random subset of pixels is extracted from the image in the RGB color space
\item Each pixel is attributed a color from the color-emotion palette according to a minimum distance between colors.
\item A list of the words present in the sample is ranked according to frequency.
\item The $k_{th}$ more frequent words are selected
\end{itemize}

Some authors \cite{Colombo99} \cite{Li2007} \cite{Agnieska10} propose the use of grading properties of color scales like HSV to perform  modulation over other color atributes, for example performing a fuzzy semantics over each emotion \cite{Li2007}. Instead, in order to have only a discrete account of color sensation we use the CIELAB color space representation as in \cite{Solli09} and \cite{Wang05} which is closer to human perception. The distance metric between two colors is given by:

\begin{equation}
\Delta E_{ab}^*=\sqrt{(L_2^*-L_1^*)^2+(a_2^*-a_1^*)^2+(b_2^*-b_1^*)^2}
\end{equation}
Where $L^*$, $a^*$ and $b^*$ are CIELAB coordinates. The color correspondence between the original colors of the image is obtained by minimization of $\Delta E_{ab}^*$ over the color-emotion palette.

\begin{figure}[H]
\begin{minipage}[b]{0.3\linewidth}
\centering
\includegraphics[width=\textwidth]{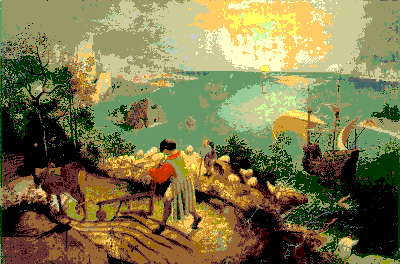}
\caption{Indexed image}
\label{fig:figure1}
\end{minipage}
\hspace{0.5cm}
\begin{minipage}[b]{0.7\linewidth}
\centering
\includegraphics[width=\textwidth]{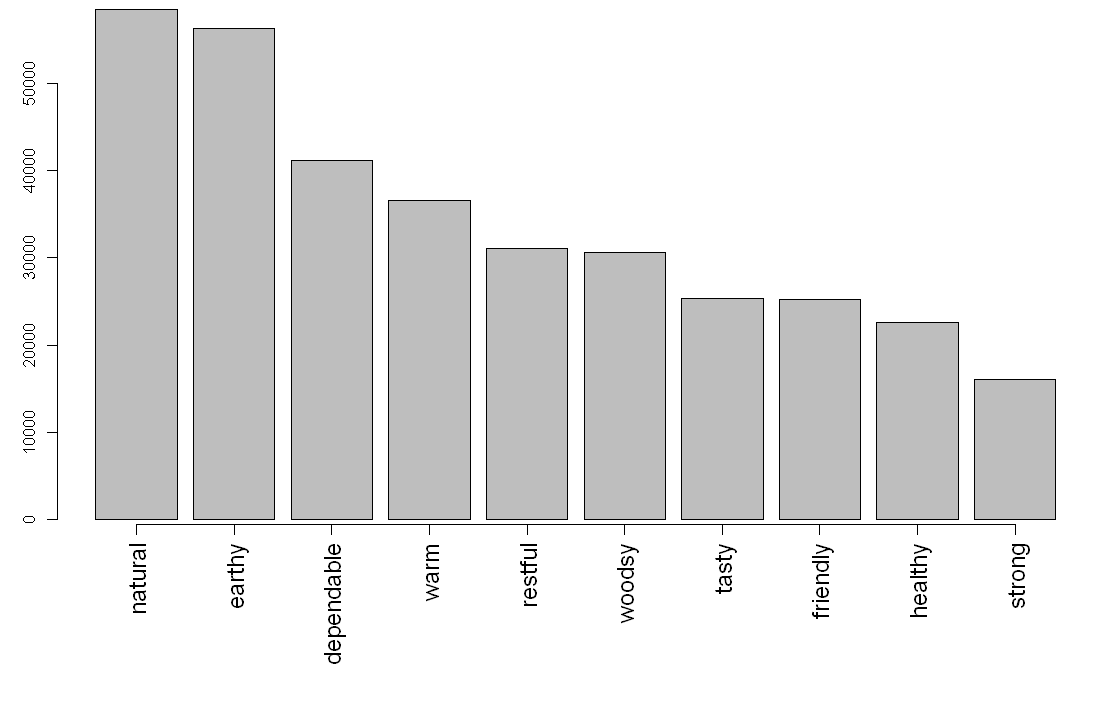}
\caption{Frequency of the 10 most frequent words associated with colors in figure \ref{fig:figure1}}
\label{fig:figure2}
\end{minipage}
\vspace{0.5cm}
\begin{minipage}[b]{0.25\linewidth}
\centering
\includegraphics[width=\textwidth]{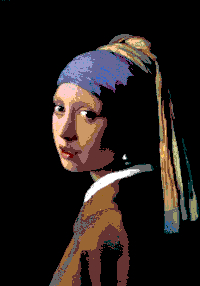}
\caption{Indexed image}
\label{fig:figure3}
\end{minipage}
\hspace{1.2cm}
\begin{minipage}[t]{0.7\linewidth}
\centering
\includegraphics[width=\textwidth]{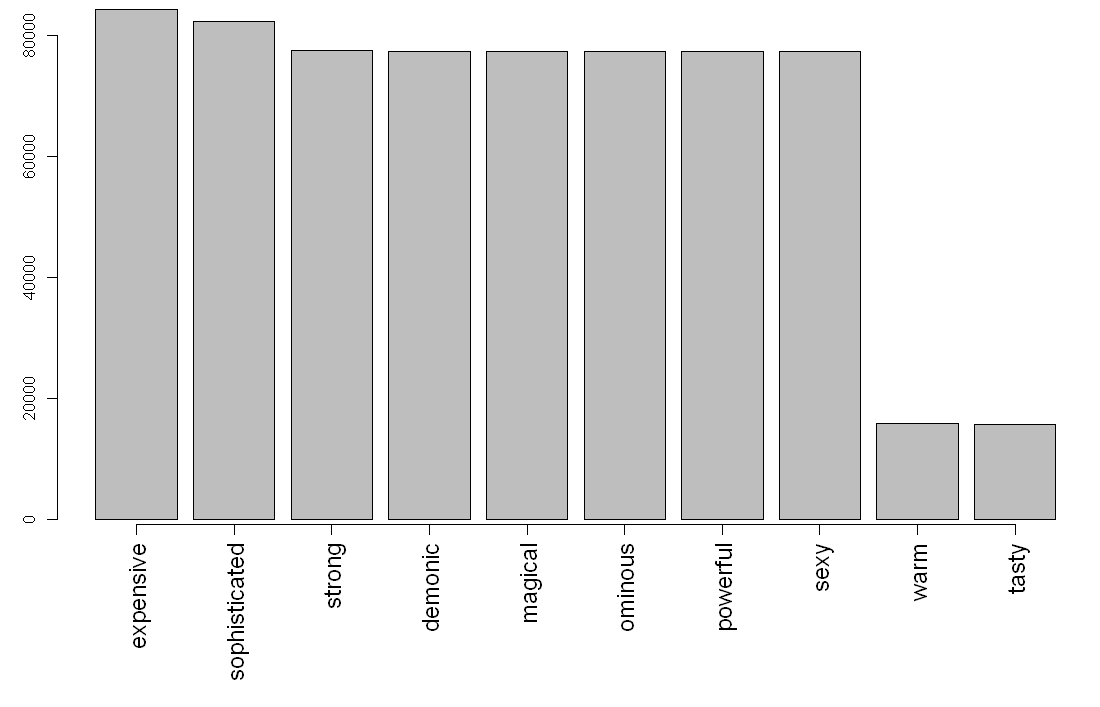}
\caption{Frequency of the 10 most frequent words associated with colors in figure \ref{fig:figure3}}
\label{fig:figure4}
\end{minipage}
\end{figure}

\begin{figure}[H]
\begin{minipage}[b]{0.5\linewidth}
\centering
\includegraphics[width=\textwidth]{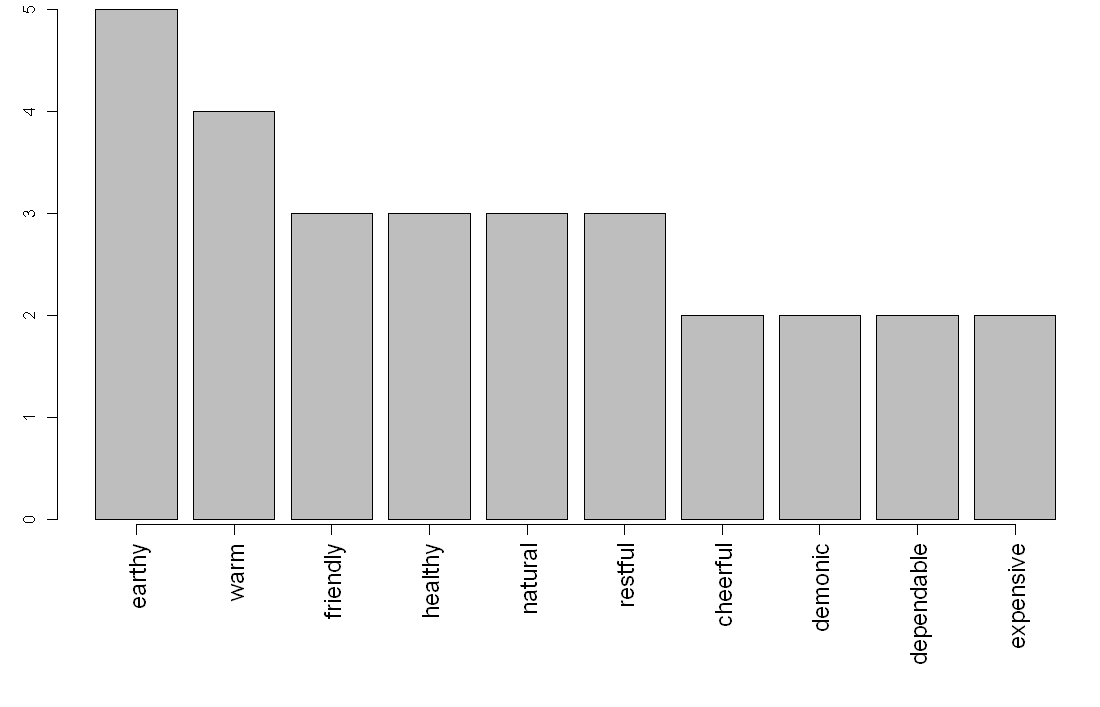}
\caption{10 most frequent words in 100 pixels sample of Figure 1.}
\label{fig:figure5}
\end{minipage}
\hspace{0.5cm}
\begin{minipage}[b]{0.5\linewidth}
\centering
\includegraphics[width=\textwidth]{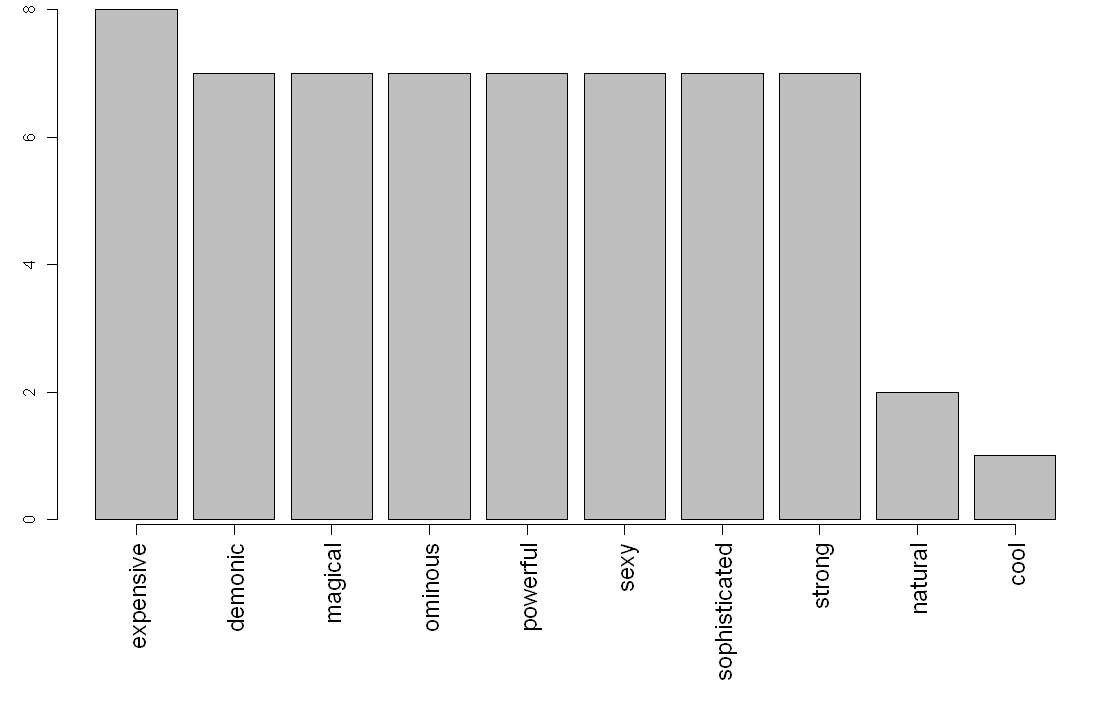}
\caption{10 most frequent words in 100 pixel sample of Figure 3.}
\label{fig:figure6}
\end{minipage}
\end{figure}

Figures \ref{fig:figure1} to \ref{fig:figure4} depict two examples of pictures with annotation. Both charts refer to a total sampling of the images. In 
Figures \ref{fig:figure5} and \ref{fig:figure6} depict the same process this time applied to a random sample of only 100 pixels from both images. We may notice that given the narrow window of the sample over the image the annotation is still approximate to the full statistics of each image. The y axis reflects the number os annotated pixels. However we may notice also that, if the picture presents big monochromatic areas it will inevitably tend to maintain the annotation even in case of a very reduced sub-sampling. This way the examination of the relation between picture complexity and information reduction becomes an important issue.

In the Appendix several other pictures are present with the corresponding histograms of colors and most frequent words.

\section{Coding}
\label{sec:code}
Sampling in the context of color \textit{inhomogeneity} as already been studied by some authors. Color entropy \cite{Rigau03}\cite{Rigau02} is a good measure of variability and can successfully aid in adaptive sampling. 

We are interested in knowing how much error is introduced in the process of information reduction. From classical information theory we know that a lower bound for the probability of error is given by Fano's inequality \cite{Fano61}. For discrete variables $X$ and $Y$ taking values on the same alphabet $\mathcal{X}=\{1,2,...\}$. In our case $\mathcal{X}$ is the finite set of RBG colors of the digital images. Now suppose we have a set  $\mathcal{Y}=\{warm,sophisticated,fruty,...\}$, of coded emotions associated with each color for which:
\begin{align}
\epsilon &= \underset{f:\mathcal{X} \mapsto \mathcal{Y}}{min} \mathbb{P}[Y \neq f(X)]\\
	&= \sum_x P_X(x)(1-\underset{y}{max}P_{Y|X}(y|x))\\
	&\leq 1-\underset{y}{max}P_Y(y)
\end{align}
where the minimum in (2) is achieved by the maximum a posteriori (MAP) estimator and (4) holds by the suboptimal choice $f(y) = \underset{x}{argmax} P_X(x)$. Then $\epsilon$ is the error introduced in the selection process by choosing a word $w \in \mathcal{Y}$ different that the word that must be selected in the original set.
If $\mathcal{X}$ is a finite set, Fano's inequality \cite{Fano61} relates the conditional entropy, also designated by equivocation, of Y given X and the error probability $\epsilon$ by
\begin{equation}
H(Y|X)\leq \epsilon \log{(| \mathcal{Y} |-1)}+h(\epsilon)
\end{equation}
where
\begin{equation}
h(x) = x \log{\frac{1}{x}}+(1-x)\log{\frac{1}{1-x}}
\end{equation}
In our case $Y$ is the discrete variable result of a mapping from colors to emotions. We may admit for performance measures  that this  random process is the result of an uniform uniform random sampling on $\mathcal{X}$ so that:
\begin{align}
p(x,y)&=p(y|x)p(x)\\
	&=k_x p(x)^2
\end{align}
Where $0\leq k_x \leq 1$. In which the perception of the emotion related to the color is directly $k_x$ proportional to the presence or absence of the color $x$ in the image.\\

Then we have:
\begin{align}
H(Y|X) &= \sum_{x\in\mathcal{X}, y \in \mathcal{Y}}p(x,y)\log{\frac{p(x)}{p(x,y)}}\\
	&= \sum_{x\in\mathcal{X}, y \in \mathcal{Y}}p(x,y)\log{\frac{1}{k_x p(x)}}\\
	&= - \sum_{x\in\mathcal{X}}p(x)\log{k_xp(x)}\\
	&=H(X)- \sum_{x\in\mathcal{X}} p(x) \log{k_x}
\end{align}
so we have that the minimum error increase by executing a set of $k_x$ choices on the proportion of color from the image, relative to an admissible optimal annotation of the image $\mathcal{X}$, is closely proportional to:
\begin{equation}
\nearrow \epsilon \sim  H(X)- E(\log{k_x})
\end{equation}
This equation seems quite reasonable. The error introduced in the emotion coding process is lower, the best case, when $ \forall x : k_x=1$, in which for every color proportion there is an associated emotion. In this case the lower bound for the error solely dependens on the entropy of the original image. In worst cases, in which $k \rightarrow 0$ the lower bound for the error would not definitively depend on the emotional coding so $\epsilon \rightarrow \infty$, but on the lack of coding.\\
In our case, which is not the proportional case but a reasoned one, we have worst than optimal correspondence between emotions and colors, in which the emotions are directly related to the proportional prevalence of colors. Given that we distinguish between 187 emotions and that the original image has a full 8bit RGB palette, each $k_x$ is approximatively given by:
\begin{align}
\forall x : k_x &\simeq 187 / 256^3\\
 &=10^{-7}\\
H(Y|X) &\simeq H(X)+16 bits
\end{align}
Which is the information introduced by the coding process when emotions are simply given by the proportion of each color. Closer proximity with the real information contained in the image, raising $k_x$ for each color and assuming all the emotional content in the image in included in its color content, given by $H(X)$, would lower this lower bound for the error.

\section{Sampling}
\label{sec:sampl}
In our essays we choose to implement a sampling without replacement process. This type of process has an associated probability which is given by a multivariate hypergeometric distribution:
\begin{equation}
p(x_i=k_i)=\frac{\Pi_{i=1}^I (^{m_i}_{k_i})}{(^N_n)}
\end{equation}
In which $k_i$ is a number of sampled pixels with the color belonging to the categories $i=\{1,2,... I\}$ with corresponding proportion in the original image given by $m_i$, from an $N$ pixels image sampled $n$ times without replacement of the pixels.

We would like to know if the process could be further simplified by executing a sampling with replacement process. This process corresponds to the simply observing of states in some pixels in the image. The equivalent distribution is a multinomial distribution with an expected value identical to the hypergeometric given by:
\begin{equation}
E(X_i)=n \frac{m_i}{N}
\end{equation}
As we are choosing the most frequent emotions, which are given by the most higher values of equation (18) between all categories $i$, the two processes are equivalent for our purposes, as both obey the same expected value equation for $X_i$. Sampling with or without replacement is equivalent. And this is evident in the samples that we've tested. So the most adequate process can be implemented without bias in the 'most relevant emotions' of the pictures selection. The ranking is equivalent to the ranking performed over the original image. Moreover, we may ask which is the resemblance between sampling without replacement and to not care at all of the pixels already examined, sampling with replacement?

This question is asked looking for the second moments of the distributions, variance and covariance between sampled pixels are given by:
\begin{align}
Var(X_i)=n\frac{m_i}{N}(1-\frac{m_i}{N})c\\
Cov(X_i,X_j)=-n\frac{m_i m_j}{N^2}c
\end{align}
with $c = \frac{N-n}{N-1}$ in the hypergeometric case, and $c=1$ in the multinomial case\\

As we make c very close to one by lowering the size of the sample, the two processes get similar. In our examples we have $c=\frac{105600-100}{105600-1}=0.999$ and $c=\frac{28147-100}{28147-1}=0.996$ as for Fig.1 and Fig. 3, respectively. So the process of executing a sampling with replacement and without replacement with small $n$ is very similar. 

We may conclude that the process of sampling and observing without much detail, with small $n$, in the case of ranking a small set of categories $I$, with random sampling with replacement of the pixels in the image is reasonable to implement.

\section{Conclusions}
\label{sec:conc}
In this paper we've presented a novel way of cataloging pictures according to emotions, thereby reducing the semantic gap between the cataloged images and the user. We've showed that the process of classification, although subjective, might be relevant according with the emotional content of the pictures sampled. We've showed also that although this process of emotion classification introduce inevitable errors of classification, this errors maybe constrained within the limits of Fano's inequality and also that the sampling process may be simplified, with speed advantage, by sampling without replacement.

\bibliographystyle{spbasic}
\bibliography{emotion_annotation_images}

\newpage
\section*{Annex}
In this Annex the emotion content according with \cite{Morton97} coding is reported for several famous pictures.



\section*{Supplementary material}

## Landscape with the Fall of Icarus (1560s) - Pieter Bruegel the Elder

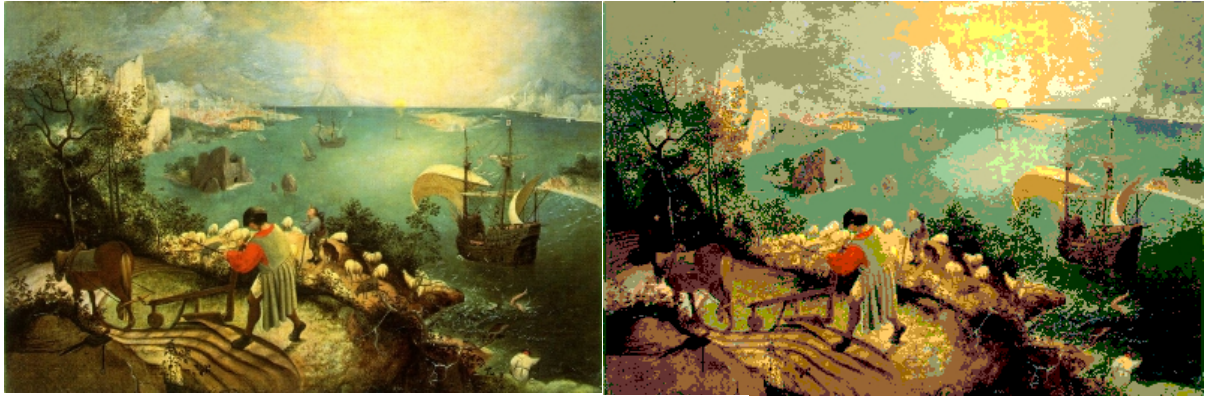

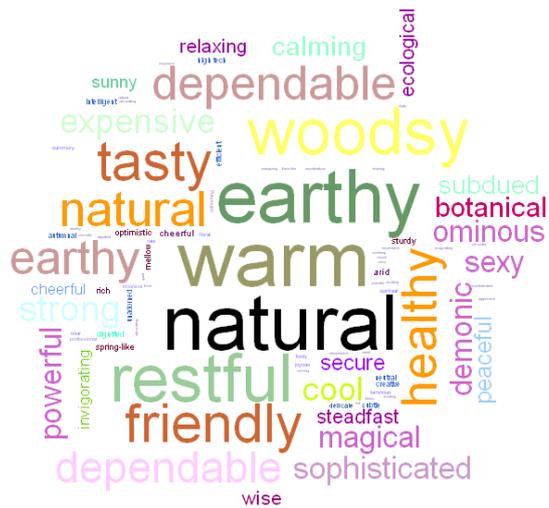

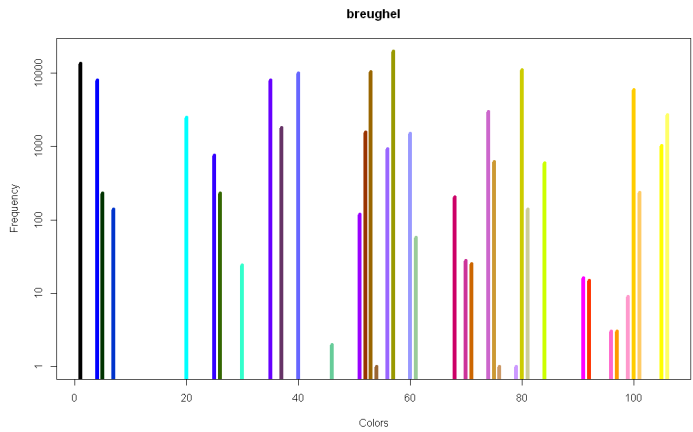

## Log histogram of colors

[1] earthy, restful
[2] powerful, sophisticated, strong, sexy, magical, demonic, ominous, expensive
[3] natural, restful, subdued
[4] dependable, friendly, tasty, healthy, earthy, natural, woodsy, warm
[5] natural, botanical, peaceful, dependable, calming, cool
[6] dependable, relaxing, friendly, natural, tasty, earthy, woodsy, warm

## Most frequent colors and emotions

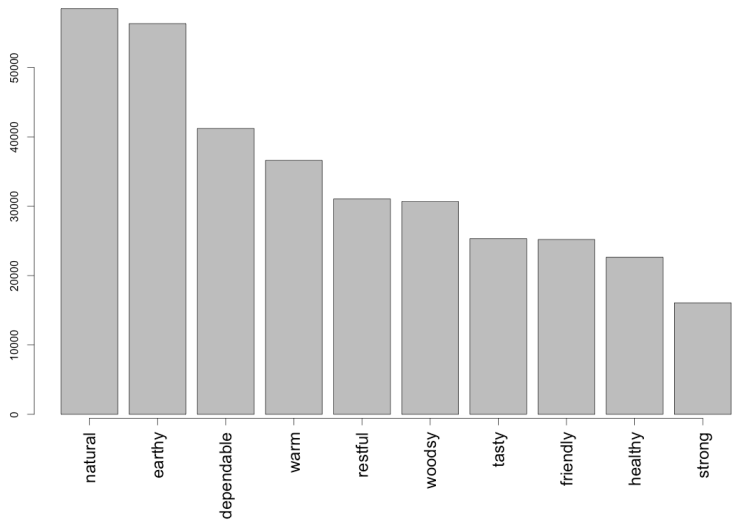

## Full pixel emotion classification

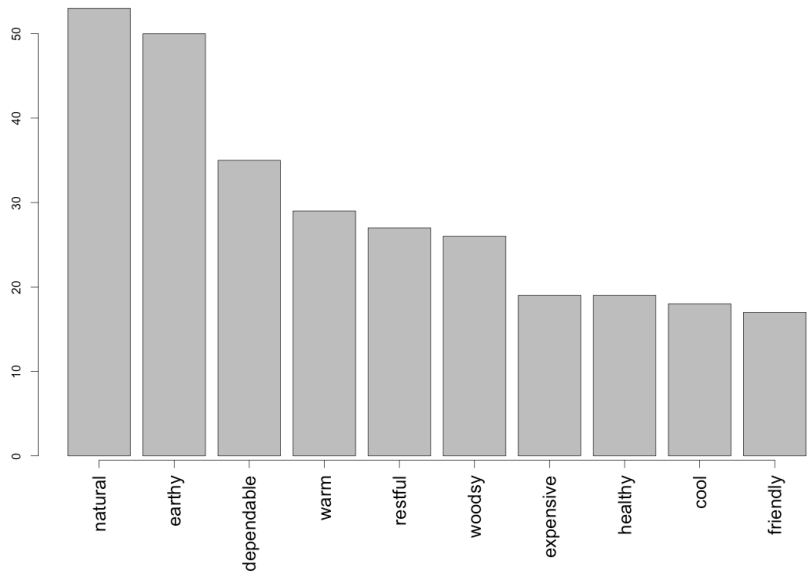

**100 pixels emotion classification**

**Untitled (1985) - Keith Haring**

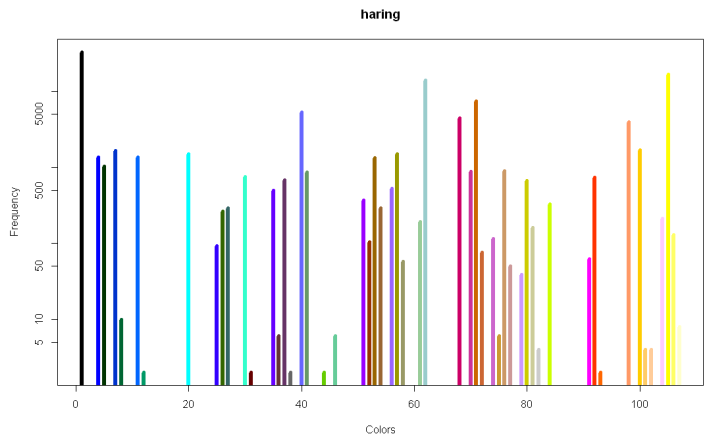

## Log histogram of colors

[1] powerful, sophisticated, strong, sexy, magical, demonic, ominous, expensive
[2] joyous, lively, spiritual, luminous, sunny, summery, floral, warm
[3] peaceful, calming, quiet, passive, ethereal, spiritual, cool
[4] zesty, spicy, natural, invigorating, tasty, healthy, earthy, autumnal, warm
[5] natural, botanical, peaceful, dependable, calming, cool
[6] invigorating, spicy, tasty, earthy, autumnal, warm

## Most frequent colors and emotions

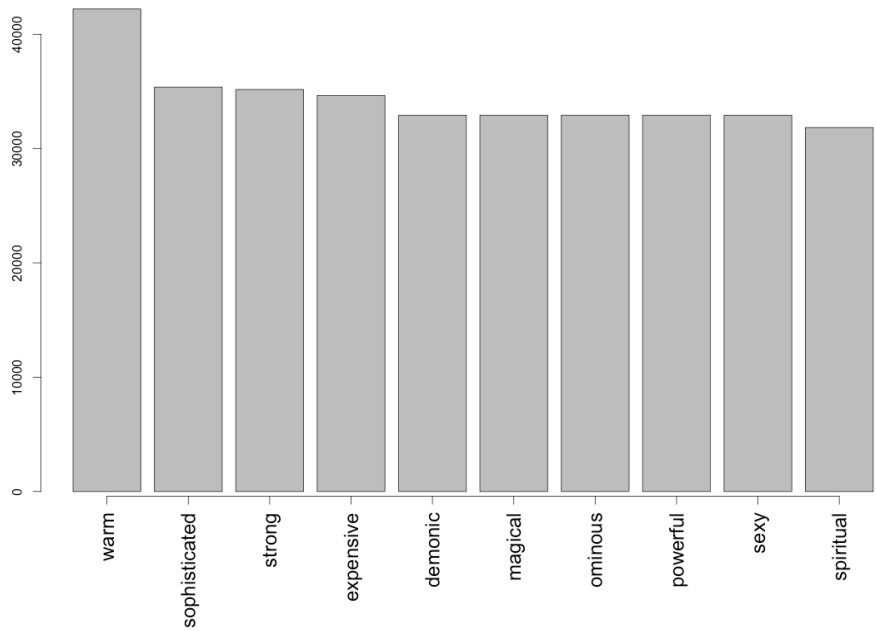

## Full pixel emotion classification

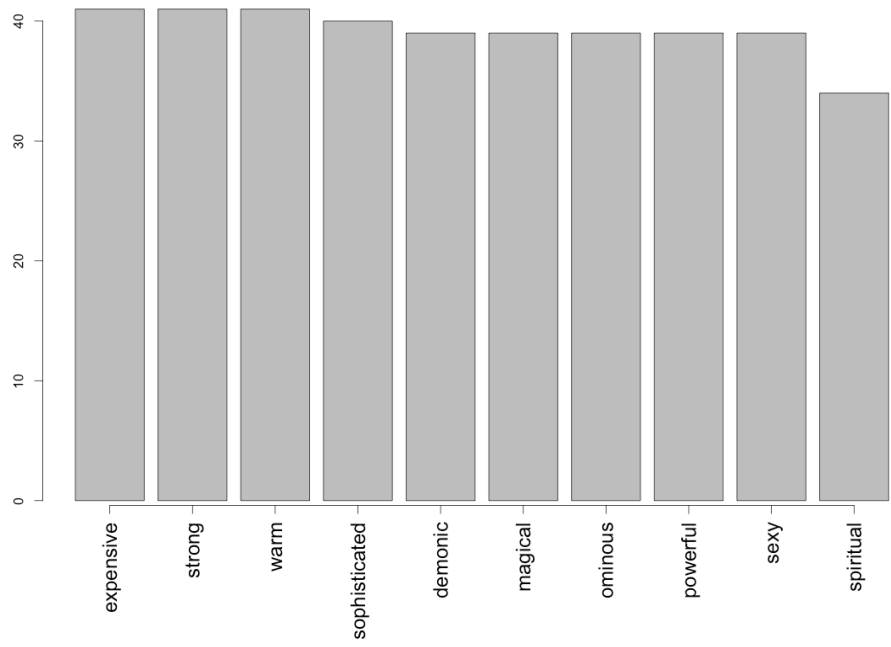

**100 pixels emotion classification**

**Les Demoiselles d'Avignon (1907) - Pablo Picasso**

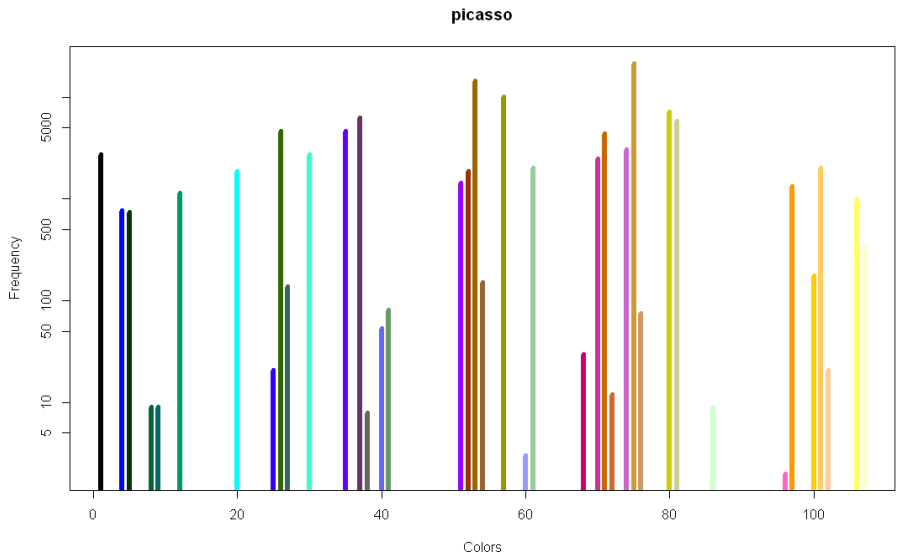

## Log histogram of colors

[1] earthy, tasty, friendly, natural, healthy, warm
[2] dependable, friendly, tasty, healthy, earthy, natural, woodsy, warm
[3] earthy, restful
[4] natural, restful, subdued
[5] dignified, intelligent, high tech, creative, efficient, expensive, unadorned, subtle, neutral, cool
[6] calming, unadorned, unpretentious, neutral, subtle, subdued, cool

## Most frequent colors and emotions

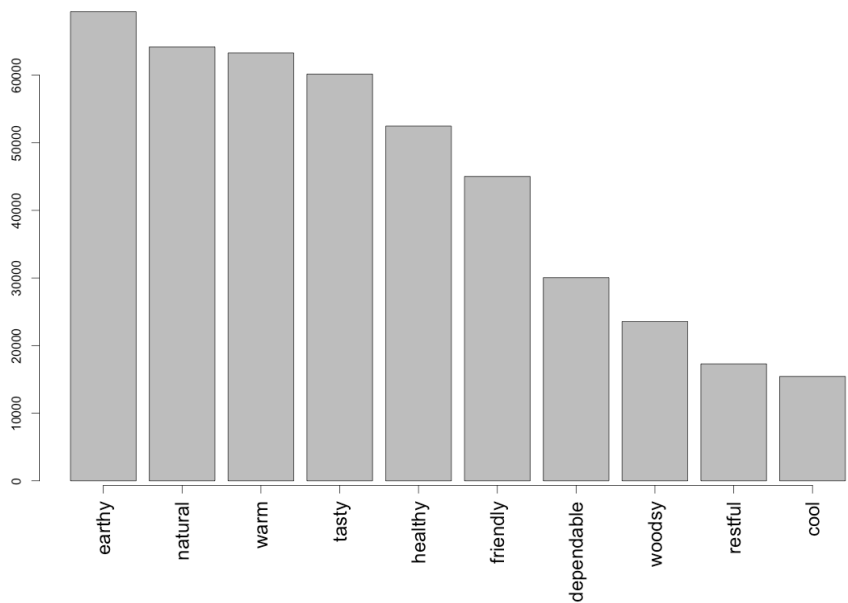

## Full pixel emotion classification

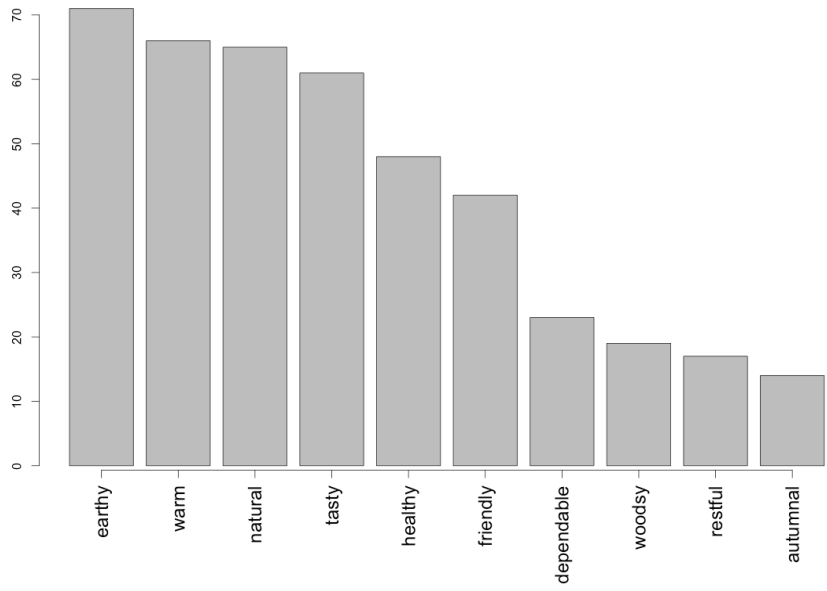

**100 pixels emotion classification**

## The Key (1946) - Jackson Pollock

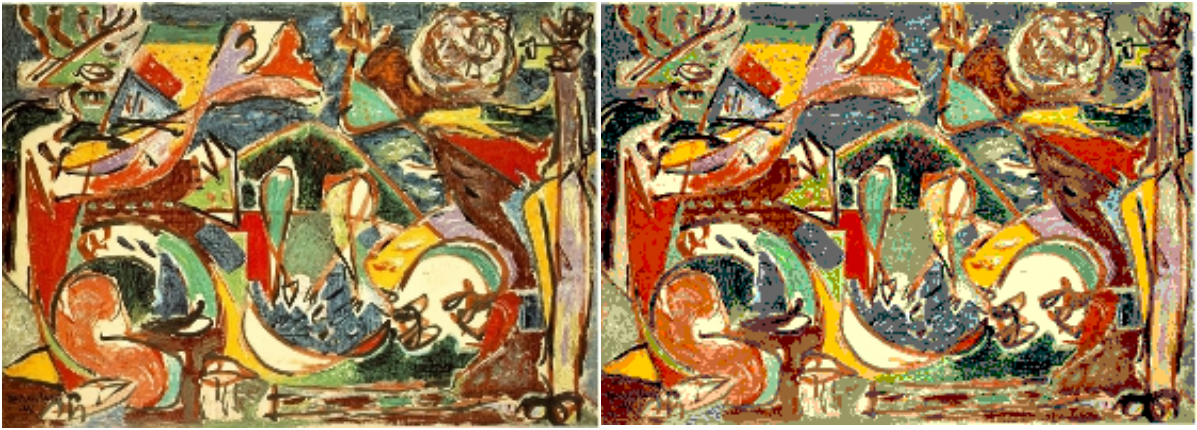

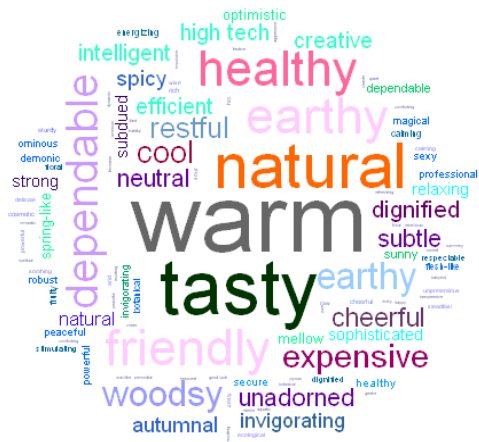

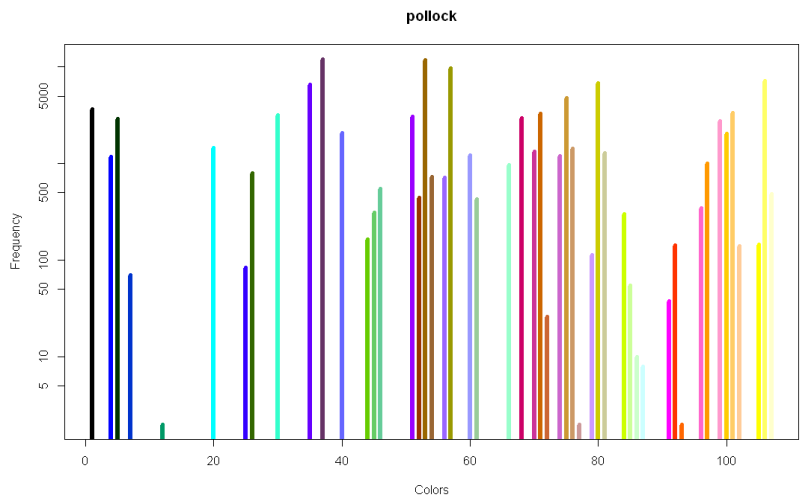

## Log histogram of colors

[1] dignified, intelligent, high tech, creative, efficient, expensive, unadorned, subtle, neutral, cool
[2] dependable, friendly, tasty, healthy, earthy, natural, woodsy, warm
[3] earthy, restful
[4] mellow, optimistic, cheerful, spring-like, warm
[5] natural, restful, subdued
[6] dependable, relaxing, friendly, natural, tasty, earthy, woodsy, warm

## Most frequent colors and emotions

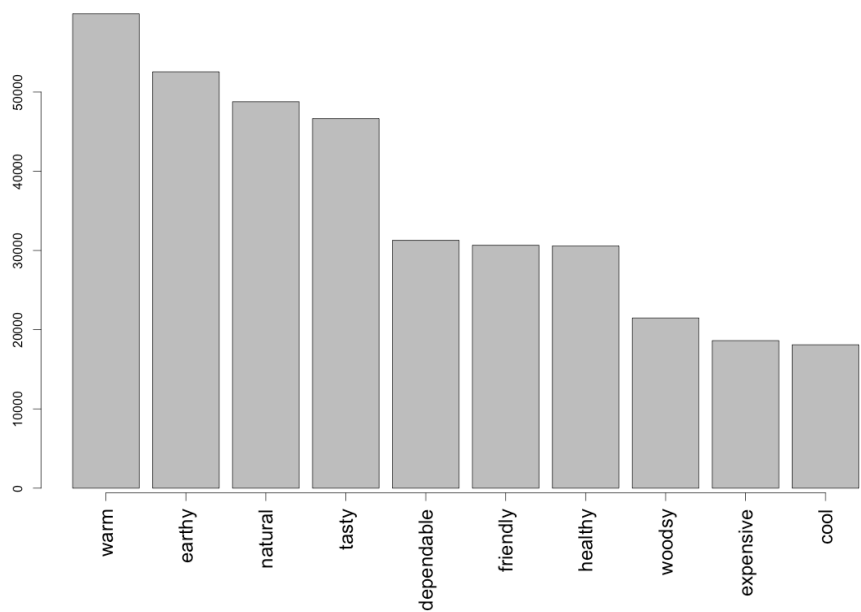

## Full pixel emotion classification

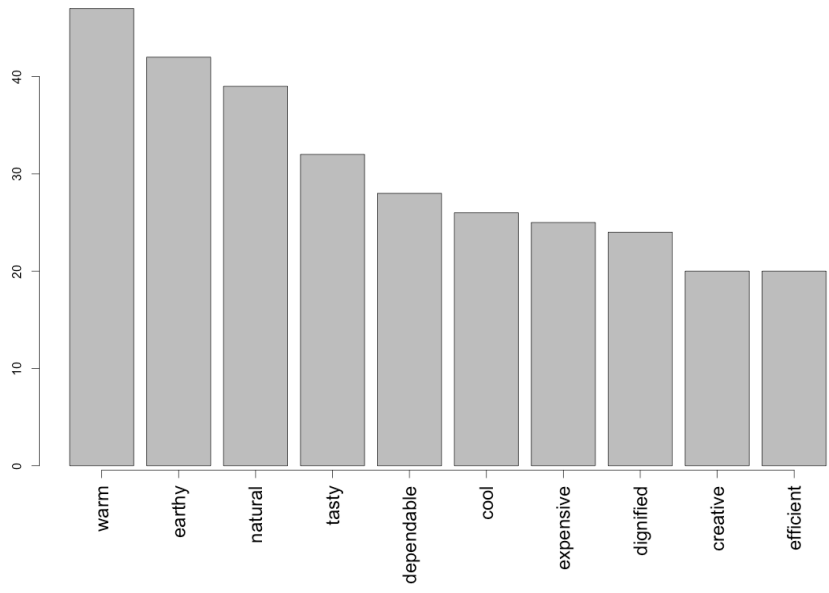

**100 pixels emotion classification**

**The Night Watch (1642) - Rembrandt**

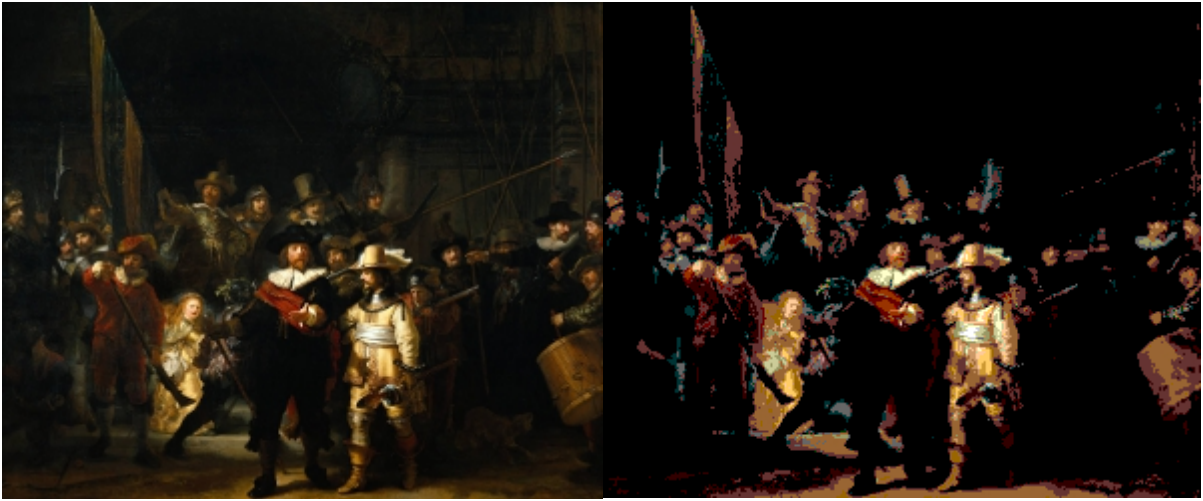

rembrand

## Log histogram of colors

[1] powerful, sophisticated, strong, sexy, magical, demonic, ominous, expensive
[2] dependable, relaxing, friendly, natural, tasty, earthy, woodsy, warm
[3] dependable, friendly, tasty, healthy, earthy, natural, woodsy, warm
[4] dependable, strong, sturdy, natural, rich, tasty, earthy, woodsy, warm
[5] dependable, dignified, professional, respectable, secure, sophisticated, expensive
[6] dignified, intelligent, high tech, creative, efficient, expensive, unadorned, subtle, neutral, cool

## Most frequent colors and emotions

## Full pixel emotion classification

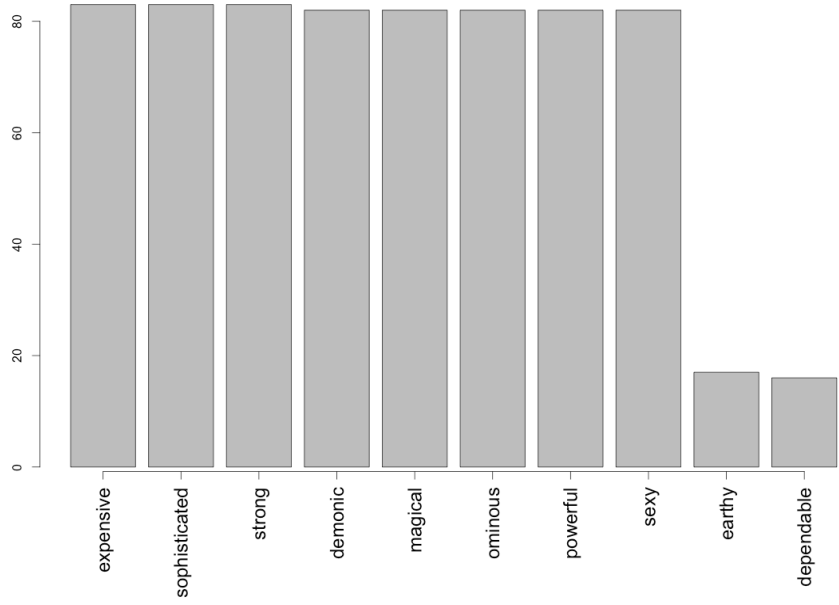

**100 pixels emotion classification**

**Portrait of Norman Rockwell Painting the Soda Jerk (1953) - Norman Rockwell**

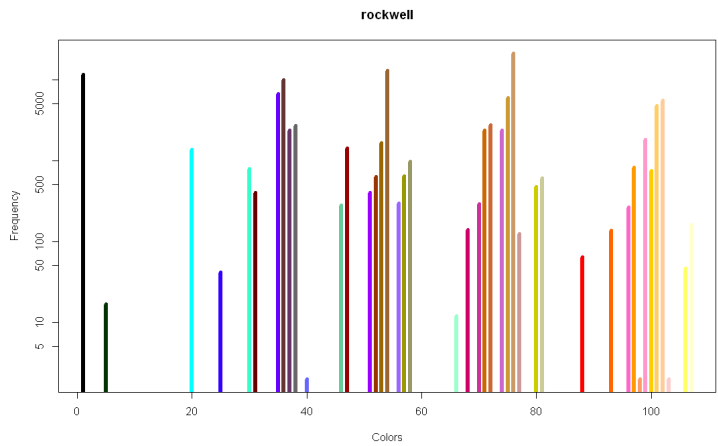

**Log histogram of colors**

[1] sophisticated, relaxing, cosmetic, warm
[2] relaxing, friendly, natural, tasty, warm
[3] powerful, sophisticated, strong, sexy, magical, demonic, ominous, expensive
[4] sophisticated, elegant, philosophical, spiritual, artistic
[5] dependable, relaxing, friendly, natural, tasty, earthy, woodsy, warm
[6] earthy, tasty, friendly, natural, healthy, warm

**Most frequent colors and emotions**

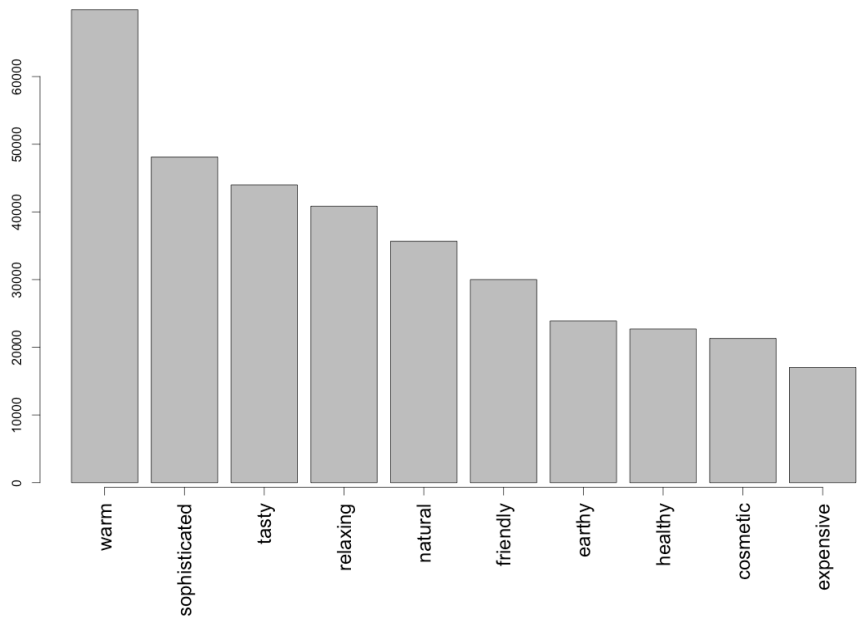

**Full pixel emotion classification**

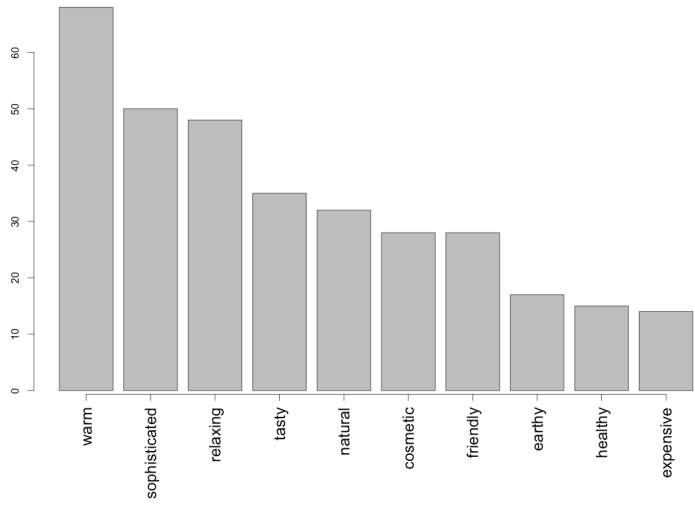

**100 pixels emotion classification**

**Blue Orange Red (1961) - Mark Rothko**

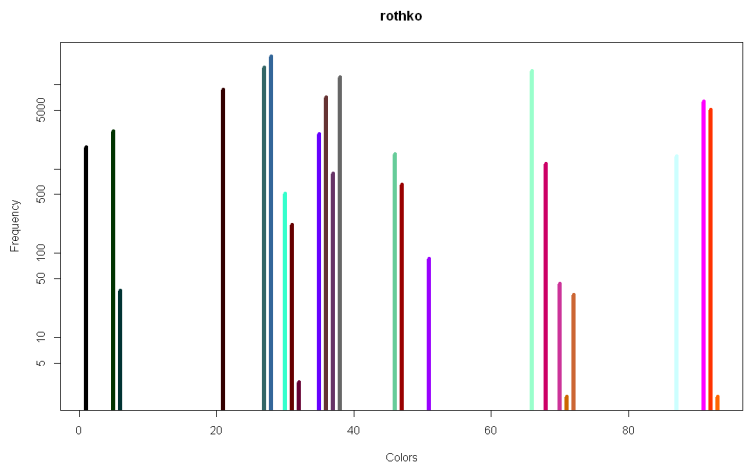

## Log histogram of colors

[1] secure, trustworthy, understanding, calming, cool
[2] dignified, dependable, professional, respectable, intelligent, peaceful
[3] invigorating, powerful, zesty, tasty, spicy, hot
[4] dignified, intellectual, respectable, expensive, sophisticated, creative
[5] regal, dignified, elegant, expensive, philosophical, intelligent, spiritual, mysterious
[6] sophisticated, elegant, philosophical, spiritual, artistic

## Most frequent colors and emotions

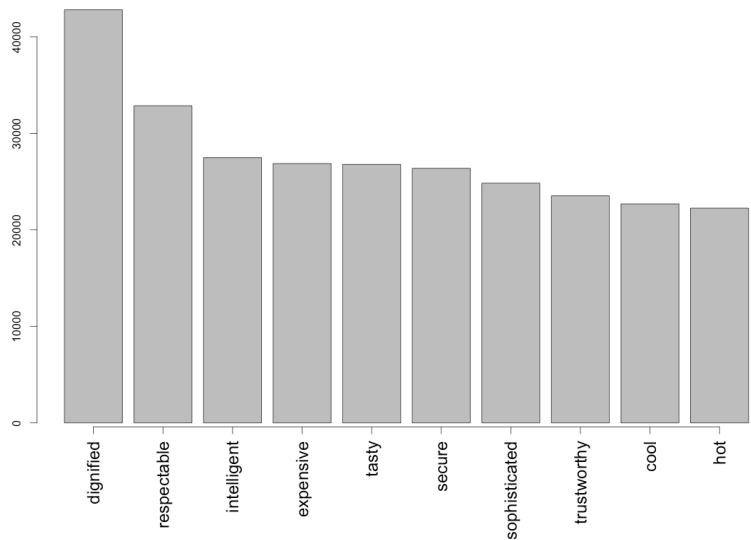

## Full pixel emotion classification

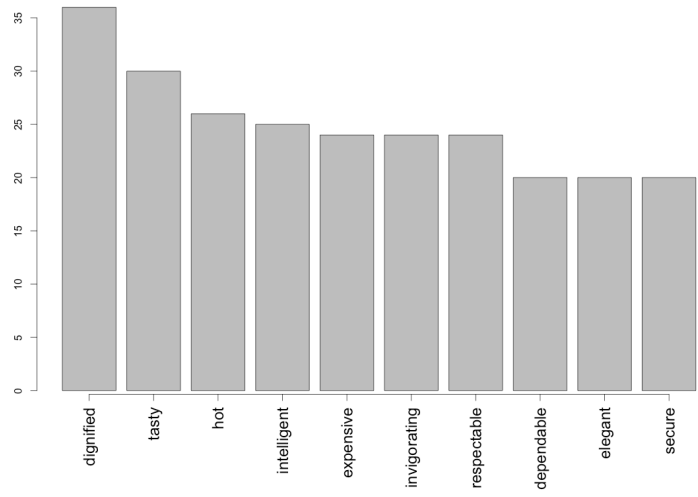

**100 pixels emotion classification**

**The Slave Ship (1781) - J. M. W. Turner**

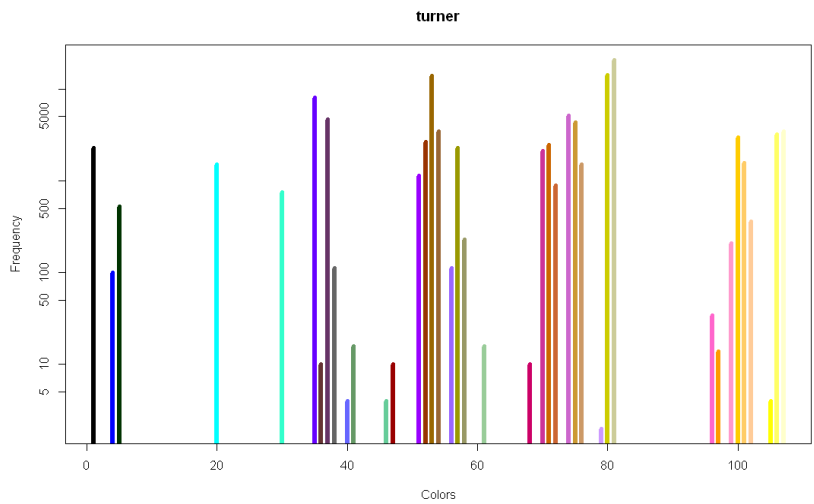

**Log histogram of colors**

[1] calming, unadorned, unpretentious, neutral, subtle, subdued, cool
[2] natural, restful, subdued
[3] dependable, friendly, tasty, healthy, earthy, natural, woodsy, warm
[4] dependable, relaxing, friendly, natural, tasty, earthy, woodsy, warm
[5] earthy, natural, healthy, tasty, arid, warm
[6] dignified, intelligent, high tech, creative, efficient, expensive, unadorned, subtle, neutral, cool

**Most frequent colors and emotions**

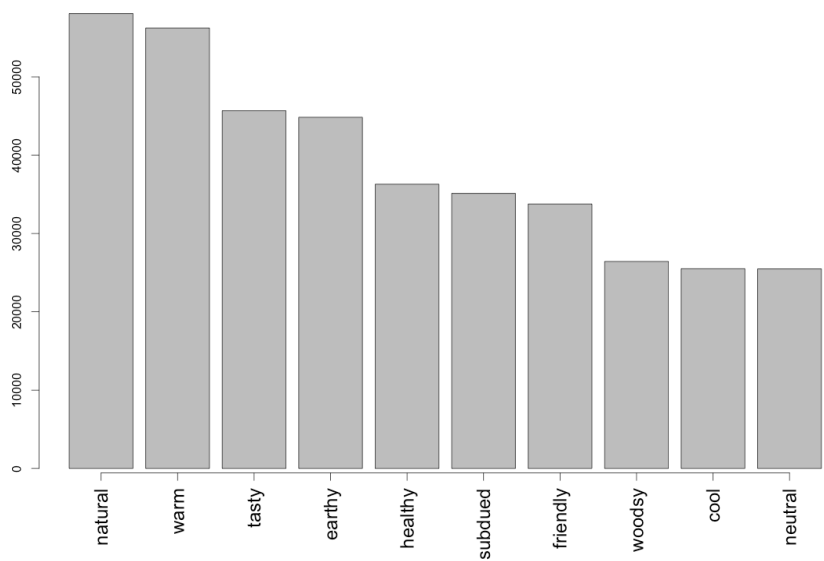

**Full pixel emotion classification**

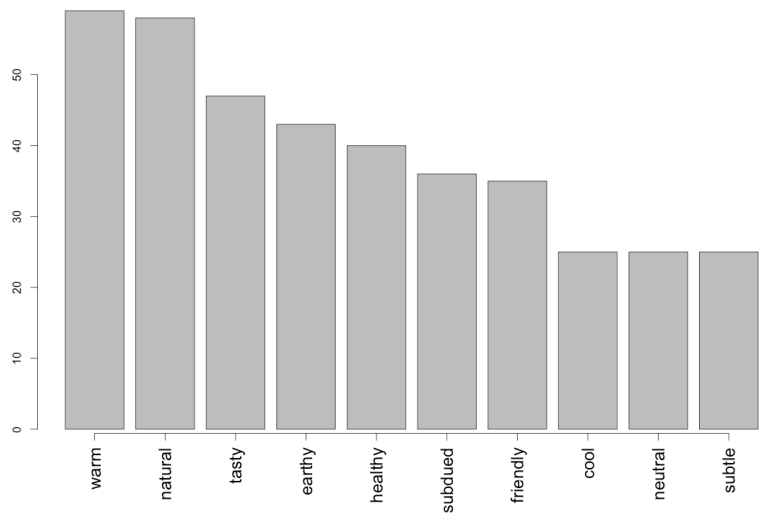

**100 pixels emotion classification**

**Bedroom in Arles (1889) - Vincent van Gogh**

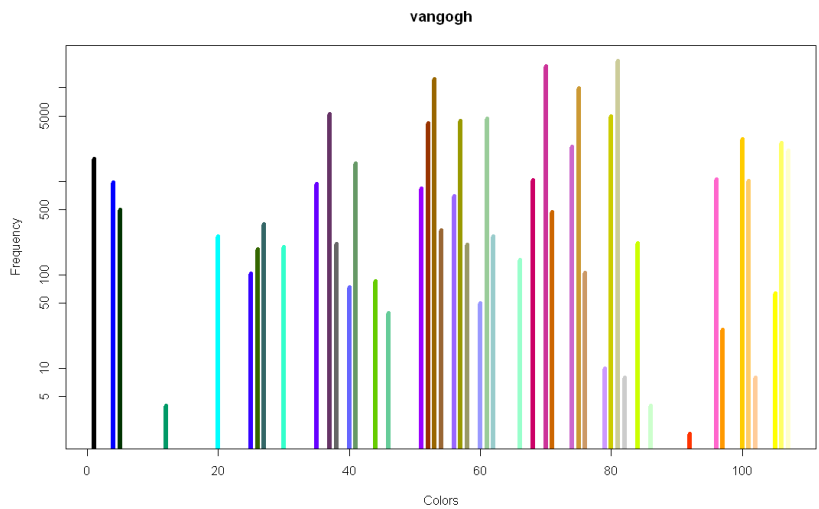

## Log histogram of colors

[1] calming, unadorned, unpretentious, neutral, subtle, subdued, cool
[2] zesty, spicy, tasty, healthy, invigorating, earthy, autumnal, warm
[3] dependable, friendly, tasty, healthy, earthy, natural, woodsy, warm
[4] earthy, tasty, friendly, natural, healthy, warm
[5] dignified, intelligent, high tech, creative, efficient, expensive, unadorned, subtle, neutral, cool
[6] natural, restful, subdued

## Most frequent colors and emotions

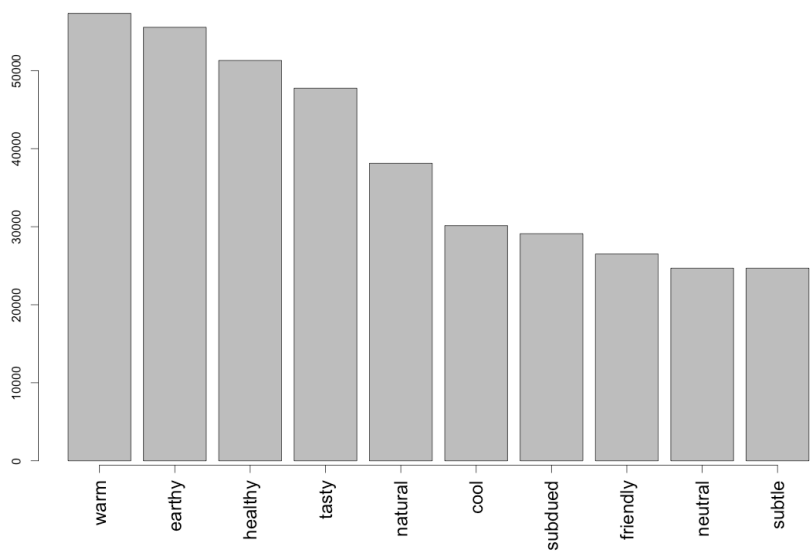

## Full pixel emotion classification

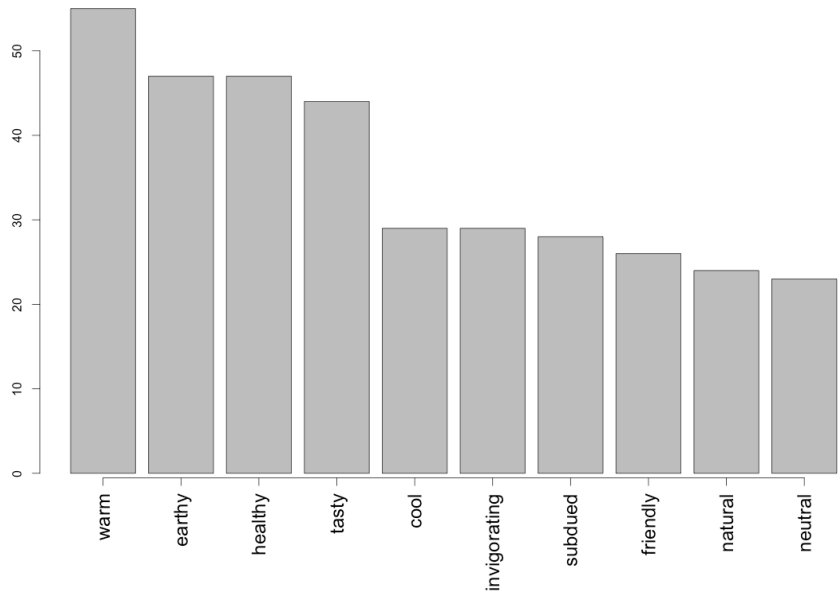

**100 pixels emotion classification**

**Girl with a Pearl Earring (1667) - Johannes Vermeer**

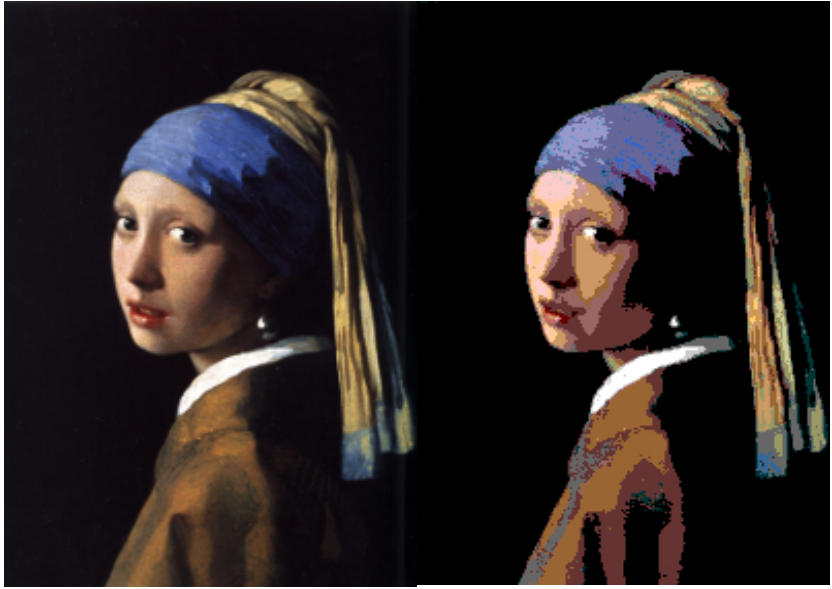

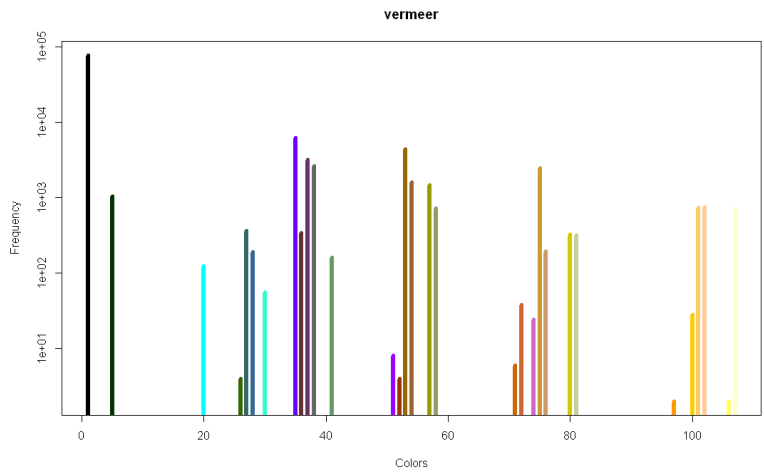

**Log histogram of colors**

1] powerful, sophisticated, strong, sexy, magical, demonic, ominous, expensive
[2] dependable, relaxing, friendly, natural, tasty, earthy, woodsy, warm
[3] dependable, friendly, tasty, healthy, earthy, natural, woodsy, warm
[4] dignified, intelligent, high tech, creative, efficient, expensive, unadorned, subtle, neutral, cool
[5] dignified, intellectual, respectable, expensive, sophisticated, creative
[6] earthy, tasty, friendly, natural, healthy, warm

**Most frequent colors and emotions**

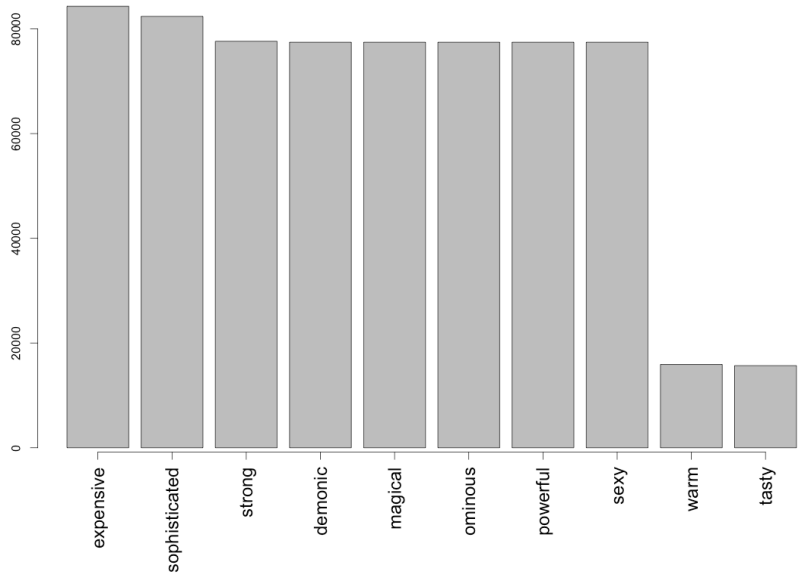

**Full pixel emotion classification**

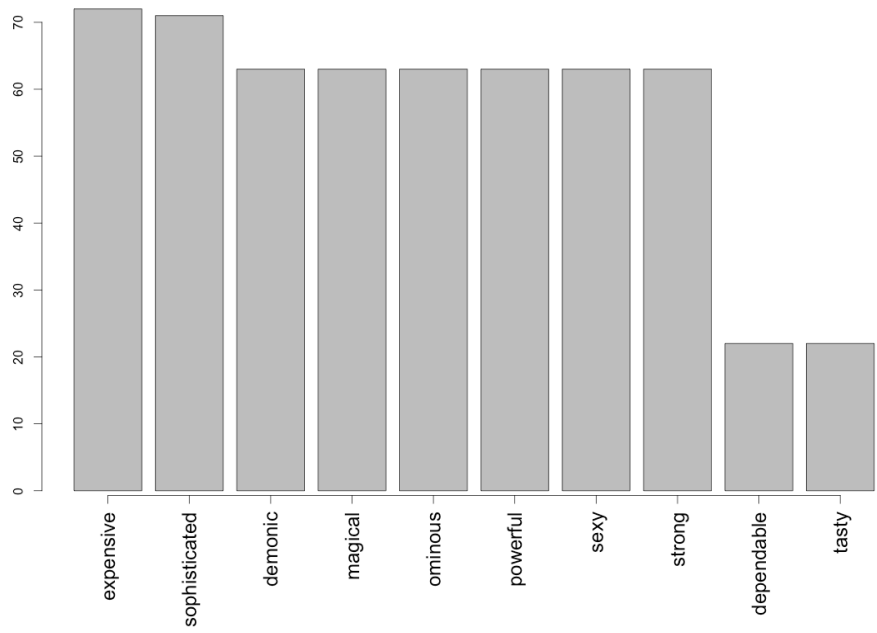

**100 pixels emotion classification**

**Campbell's Soup Cans (1962) - Andy Warhol**

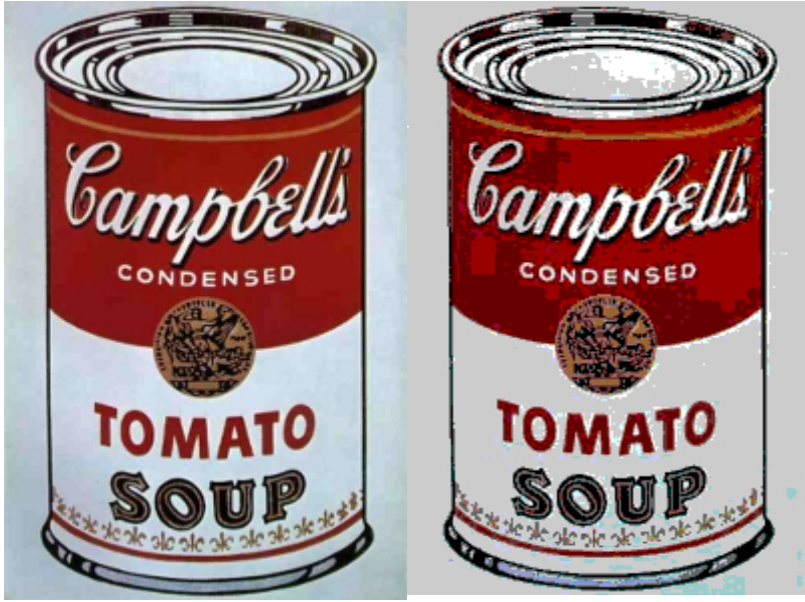

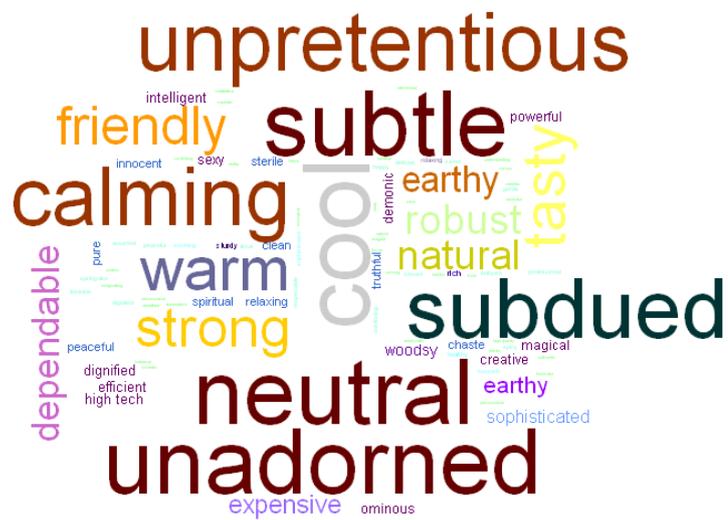

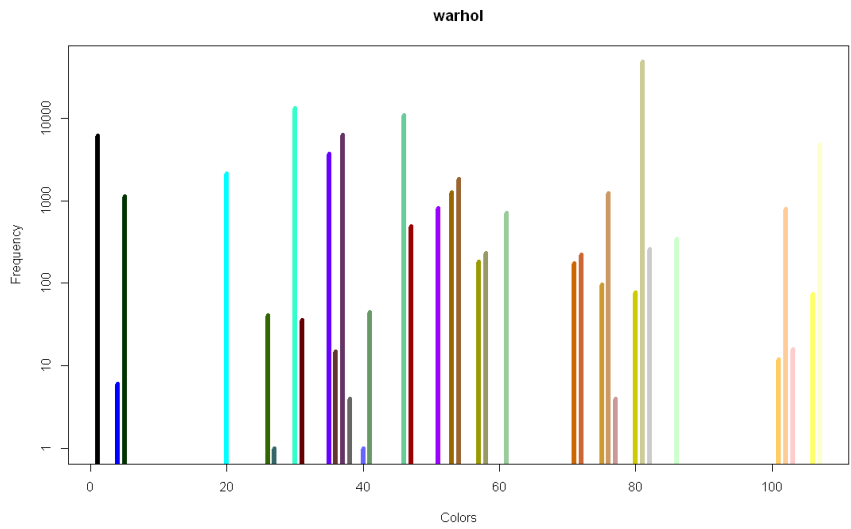

## Log histogram of colors

[1] calming, unadorned, unpretentious, neutral, subtle, subdued, cool
[2] dependable, strong, robust, friendly, natural, earthy, tasty, warm
[3] earthy, friendly, robust, strong, tasty, warm
[4] dignified, intelligent, high tech, creative, efficient, expensive, unadorned, subtle, neutral, cool
[5] powerful, sophisticated, strong, sexy, magical, demonic, ominous, expensive
[6] pure, spiritual, clean, sterile, truthful, chaste, innocent, peaceful

## Most frequent colors and emotions

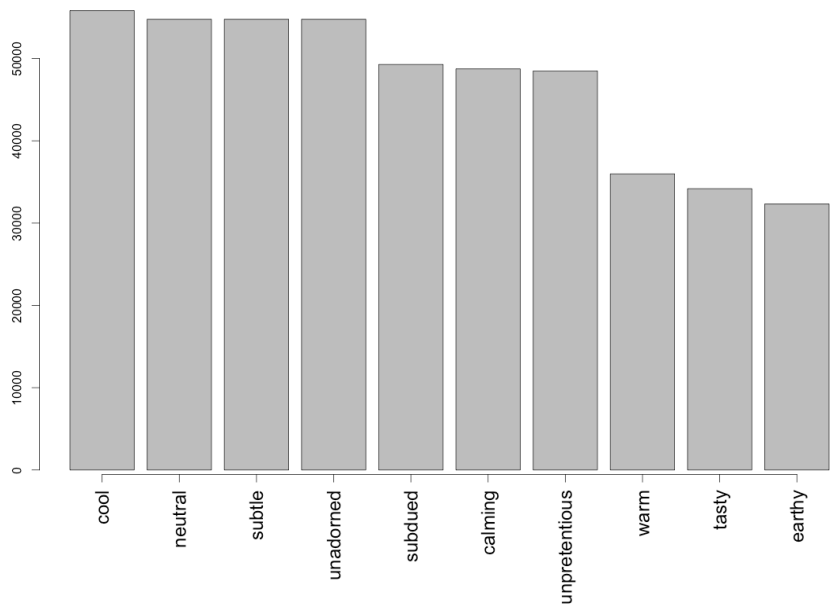

## Full pixel emotion classification

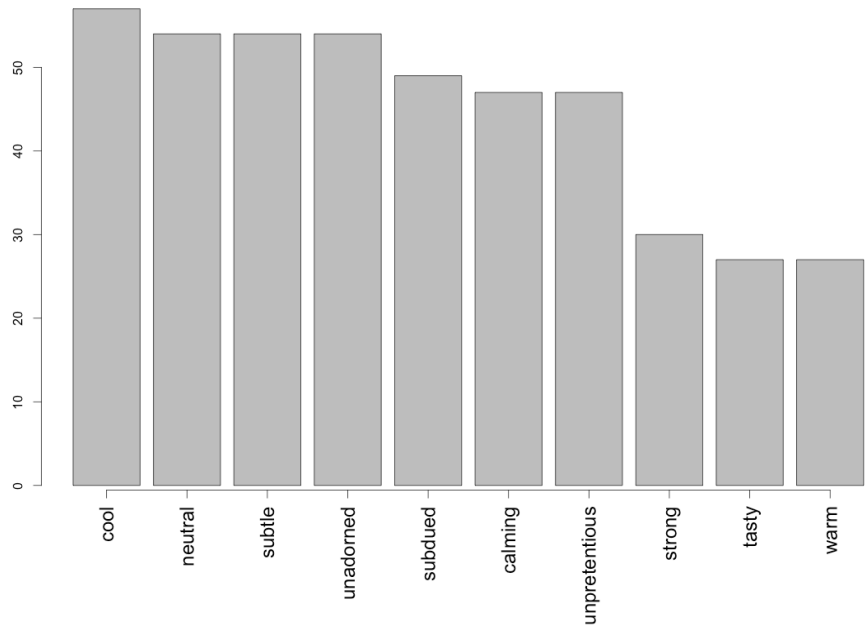

**100 pixels emotion classification**



\end{document}